\documentclass{eptcs}

\usepackage{breakurl}
\usepackage{times}
\usepackage{amssymb}
\usepackage{amsmath}
\usepackage{amsthm}
\usepackage{latexsym}
\usepackage{graphicx}

\newtheorem{definition}{Definition}
\newtheorem{lemma}{Lemma}
\newtheorem{example}{Example}
\newcommand{\Nb}{{\textbf{Nb}}}
\renewcommand{\S}{{\textbf{S}}}
\renewcommand{\L}{{\textbf{Loc}}}

\newcommand{\s}{{\mathbf{s}}}
\newcommand{\op}{{\mathbf{op}}}
\newcommand{\nby}[1]{\stackrel{ #1 }{\longrightarrow}_n}
\newcommand{\pby}[1]{\stackrel{ #1 }{\longrightarrow}_p}
\newcommand{\by}[1]{\stackrel{ #1 }{\longrightarrow}}
\newcommand{\C}{{\textbf{Ch}}}
\newcommand{\myloc}{{\sf myloc}}
\newcommand{\axiom}[1]{({\sf #1})}
\newcommand{\nil}{{\bf 0}}
\newcommand{\cond}{{\sf cond}\;}
\newcommand{\loc}[1]{{:}[\![{ #1 }]\!]}
\newcommand{\hide}{\backslash}
\newcommand{\calC}{{\mathcal C}}
\newcommand{\eqdef}{\stackrel{{\rm def}}{=}}
\newcommand{\ol}{\overline}
\newcommand{\true}{{\sf true}}
\newcommand{\sat}{{\models}}
\newcommand{\val}{{\sf val}\;}
\newcommand{\num}{{\sf num}}
\renewcommand{\a}{\alpha}
\newcommand{\fracc}[2]{\mbox{$\frac{\displaystyle{#1}}{\displaystyle{#2}}$}}

\title{A Process Calculus for Spatially-explicit Ecological Models}

\author{Margarita Antonaki\qquad\qquad Anna Philippou
\institute{Department of Computer Science\\ University of Cyprus}\\
 \email{cs05ma@cs.ucy.ac.cy \quad\qquad annap@cs.ucy.ac.cy} }
\def\titlerunning{A process calculus for spatially-explicit ecological models}
\def\authorrunning{M. Antonaki and A. Philippou}

\begin{document}
\maketitle
%\pagestyle{plain}
%\pagenumbering{arabic}
%\setcounter{page}{1}

\begin{abstract}
We propose PALPS, a Process Algebra with Locations for Population Systems.
PALPS allows us to produce spatially-explicit, individual-based models and
to reason about their behavior. Our calculus has two levels: at the first
level we may define the behavior of an individual of a population while,
at the second level, we may specify a system as the collection of individuals
of various species located in space, moving through their life cycle while
changing their location, if they so wish, and interacting with each other
in various ways such as preying on each other. Furthermore, we propose a
probabilistic temporal logic for reasoning about the behavior of PALPS
processes. We illustrate our framework via models of dispersal in
metapopulations.
\end{abstract}

\section{Introduction}

During the last decade we have witnessed an increasing trend towards
the use of formal frameworks for reasoning about biological as well
as ecological systems including process algebras
\cite{Tofts94,BIO-A,Cardelli-BC,MNS08,BIO-Pepa},
Membrane Systems~\cite{MC,CCMPPS09} and cellular automata~\cite{FM04}.
Process algebras, first proposed in
\cite{Milner-CCS,Hoare-CSP} to aid the understanding and reasoning
about communication and concurrency, provide a number of
features that make them suitable for capturing biological
processes. In particular, process algebras are especially suited
towards the so-called ``individual-based'' approach of modeling
populations, as they enable one to describe the evolution of each
individual of the population as a process and, subsequently, to
compose a set of individuals (as well as their environment) into a
complete ecological system. Features such as time, probability and
stochastic behavior, which have been extensively studied within the
context of process algebras, can be exploited to provide more
accurate models, while associated analysis tools can be used to
analyze and predict their behavior.

In this work, our aim is to introduce a process-algebraic framework
to enable spatially-explicit modeling of ecological systems. Such modeling~\cite{SE,SEIBM}
has been of special interest to conservation scientists and practitioners who have employed it
in order to predict how species will respond to specific management
schemes and guide the selection of reservation sites and reintroduction efforts, e.g.
\cite{one,two}. The use of spatially-explicit, individual-based modeling requires the
description of the environment and the individuals residing in it, including
a description of each individual's interaction with other individuals
as well as with the environment. As far as the environment is concerned, these models
typically involve the use of \emph{patches} or a \emph{lattice} to represent the habitat.
Individuals are then placed on specific locations of the modeled landscape
and their behavior, including events such as birth, mortality, and dispersal,
is simulated at the individual or the population level and analyzed.

In order to capture this type of behavior our process algebra, PALPS,
associates processes  with information about their location and
their species. The habitat is defined as a graph consisting of a
set of locations and a neighborhood relation.
Movement of located processes is then modeled as the change in the location of a process,
with the restriction that the originating and the destination locations are
neighboring locations. In addition to moving between locations,
located processes may communicate with each other by
exchanging messages upon channels. Communication
may take place only between processes which reside at the same location while special
channels allow processes to engage in preying and reproduction. Furthermore,
PALPS may model probabilistic events, with the aid of a probabilistic choice operator,
and uses a discrete treatment of time. Finally, in PALPS, each location may be
associated with a set of attributes capturing relevant information such as the capacity or
the quality of the location.
These attributes form the basis of a set
of expressions that refer to the state of the environment
and are employed within models to enable the enunciation of location-dependent behavior.

The operational semantics of our calculus is given in terms of a
labeled transition system on which we may check properties expressed
in an instantiation of the PCTL temporal logic.  We illustrate the expressiveness
of PALPS by constructing spatially-explicit individual-based
models for metapopulation dispersal.

There exists a variety of previous proposals which introduce locations or compartments
into formal frameworks, e.g.
\cite{BMMP09,Cardelli-CiE2010,Bio-PEP,Pardini-thesis,JEU08,Meta,WC-2011}, while work has been carried
out to employ these frameworks for modeling and analyzing population systems~\cite{three}.
PALPS %is to a large extent influenced by such works. However, it
departs from these works in that
it is the first process-algebraic framework
developed specifically for reasoning about ecological models as well as in its
treatment of a state and its capability of expressing state-dependent behavior. In particular,
it can be considered as an extension of WSCCS of~\cite{Tofts94} with  locations
and location attributes, while it shares a similar treatment of locations with process
algebras developed for reasoning about mobile ad hoc networks, e.g.~\cite{KP11, Galpin09}. As such,
PALPS considers a two-dimensional space where locations and their interconnections are modeled
as a graph upon which individuals may move as computation proceeds. The main feature
that distinguishes PALPS from existing formal frameworks is the fact that it associates locations
with a set of attributes that model special characteristics of locations which may be of interest
when modeling a system and the ability to express behavior of individuals that is conditional
on the values of these attributes. Examples of attributes that can be observed by individuals is
the number of individuals a location can support as well the current number of individuals
present at a location.

In the remainder of the paper we present the syntax and the
semantics of PALPS in Section \ref{proccalc}, while in Section
\ref{exampl} we provide models of metapopulation dispersal. In
Section \ref{remarks} we conclude with a discussion on future work.

\section{The Process Calculus}\label{proccalc}

In our calculus, PALPS (Process Algebra with Locations for
Population Systems), we consider a system as a set of individuals
operating in space, each possessing a species and a location
identifier. Movement in the calculus is modeled via a specialized
action whose effect is to change the location of an individual, with
the restriction that the originating and the destination locations
are neighboring locations. The notion of neighborhood is implemented
via a relation $\Nb$ where $(\ell,\ell')\in \Nb$ exactly when
locations $\ell$ and $\ell'$ are neighbors. We also use $\Nb$ as a function
and write $\Nb(\ell)$ for the set of all neighbors of $\ell$.

\subsection{The Syntax}

We continue to formalize the syntax of PALPS.  We begin by
describing the basic entities of the calculus.
\begin{itemize}
\item We assume a set of special labels $\S$ corresponding to the species
under consideration, ranged over by $\s$, $\s'$.
\item Furthermore, we assume a set of
channels $\C$, ranged over by lower-case strings. This set contains
the special channels $rep_\s$ and $prey_\s$, $\s\in\S$, which are channels used
to model reproduction of species $\s$ and preying on species $\s$.
\item Finally, we assume a set of
locations $\L$ ranged over by $\ell$, $\ell'$.  Locations can be associated
with a set of attributes that model special characteristics of locations
of interest within a system. We write $\psi$ for attributes and
$\psi_{\ell}$ for the value of attribute $\psi$ at location $\ell$.
\end{itemize}

Our calculus also employs two sets of expressions: logical expressions ranged over
by $e$ and arithmetic expressions, ranged over by $w$.
One of our main aims being to facilitate reasoning about
spatially-dependent behavior, these expressions are intended to capture
environmental (location-relevant) situations which may affect the
behavior of individuals.
Expressions $e$ and $w$, are constructed as
follows:
\begin{eqnarray*}
e & ::=& true \:\:|\:\: \lnot e \:\:|\:\: e_1 \land e_2 \:\:|\:\:w \bowtie c\\
w & ::=& c \:\:|\:\: \psi@{\ell^\star} \:\:|\:\: \s@ \ell^\star\:\:|\:\: @ \ell^\star \:\:|\:\: \op_1(w) \:\:|\:\: \op_2(w_1,w_2)
\end{eqnarray*}
where $c$ is a real number, $\bowtie\in\{=, \leq, \geq\}$ and $\ell^\star\in \L\cup\{\myloc\}$.
Let us informally consider the introduced expressions.  To begin with, logical expressions are built using the propositional calculus connectives as well as comparisons between an arithmetic expression $w$ and a constant $c$, i,e. $w\bowtie c$. Moving on to arithmetic expressions, these include
three special expressions interpreted as follows: Expression $\psi@\ell^\star$ is equal to the value of attribute $\psi$ at location $\ell^\star$. Expression $(\s @ \ell^\star)$ is equal to the number of individuals of species $\s$ at location
$\ell^\star$ and expression $@
\ell^\star$ denotes  the total number of individuals of
all species at location $\ell^{\star}$. As specified above, $\ell^{\star}$ can be an arbitrary location or the special location $\myloc$. This
 label is employed to bestow individuals
the ability to express conditions on the status of their current
location no matter where that might be as computation proceeds.
Specifically, $\myloc$  refers to the actual location of the
individual in which the expression appears and it is instantiated to
this location when the condition needs to be evaluated (see rule
\axiom{Cond} in Table~\ref{op-sem-nodes}).

Thus, arithmetic expressions are the set of all expressions formed
by arbitrary constants $c$, quantities $\psi@{\ell^\star}$, $\s @
\ell^\star$, $@ \ell^\star$and the usual unary and binary arithmetic
operations ($\op_1$ and $\op_2$) on the real numbers. Logical
expressions and arithmetic expressions are evaluated within a system
environment. The precise definition of the evaluation function is
postponed to Tables~\ref{evals} and~\ref{models}.

We may now move on to the syntax of PALPS which is given
at three levels: (1) the individual
level, ranged over by $P$, (2) the species level, ranged over by
$R$, and (3) the system level, ranged over by $S$. Their syntax is
defined via the following BNF's:
\begin{eqnarray*}
  P &::=& \nil \:\:|\:\: \eta.P\:\:|\:\: \sum_{i\in I} w_i: P_i %\: + \:P_2\:\:|\:\:P_1\: +_p \:P_2
  \:\:|\:\:
  \  \cond(e_1\rhd P_1,\ldots, e_n\rhd P_n) \:\:|\:\: C\\
  R & ::= & !rep.P\\\vspace{.1in}
  S &::=& \nil  \:\:\:|\:\:\: P\loc{\s, \ell}\:\:\:|\:\:\: R\loc{\s} \:\:\:|\:\:\: S_1
  \,|\,S_2 \:\:\:|\:\:\: S\hide L
\end{eqnarray*}
where $a\in\C$, $L\subseteq \C$, $C$ ranges over a set of process
constants $\calC$, each with an associated definition of the form
$C\eqdef P$, where the node $P$ may contain occurrences of $C$, as
well as other constants,  and
\[ \eta ::=  a \:\:|\:\: \ol{a} \:\:|\:\: %move \:\:|\:\:
           go\; \ell \:\:|\:\: \surd\,.\]
%
%and $L$ is a set of channels.

Beginning with the \emph{individual} level $P$,  process $\nil$
represents the inactive individual, that is,
an individual who has ceased to exist. $\eta.P$ describes the
individual who first engages in activity $\eta$ and then behaves as
$P$. Activity $\eta$ can be an input action on a channel $a$,
written simply as $a$, a complementary output action on a channel
$a$, written as $\ol{a}$, a movement action with destination $\ell$, $go\, \ell$,
 or the time-passing action, $\surd$. Actions of the
form $a$, and $\ol{a}$, $a\in\C$, are used to model arbitrary
activities performed by an individual e.g. eating, preying, observing the
environment as well as reproduction. Thus, for example, the actions $prey_\s$ and $\ol{prey_\s}$
are executed, respectively, by a prey of population $\s$ and a predator who is preying on individuals of population $\s$.
The tick action $\surd$ measures a
tick on a global clock and is used to separate the phases/rounds of
an individual's behavior. Essentially, the intention
is that in any given time unit all individuals perform their
available actions possibly synchronizing as necessary until they
synchronize on their next $\surd$ action and proceed to their next
round.

$\sum_{i\in I} w_i: P_i$
represents the probabilistic choice between processes $P_i$, $i\in I$.
Each alternative is associated with a probability of appearance, which is the
value to which the expression $w_i$ evaluates. The conditional
process $\cond(e_1\rhd P_1,\ldots, e_n\rhd P_n)$ presents the
conditional choice between a set of processes: it behaves as $P_i$,
where $i$ is the smallest integer for which $e_i$ evaluates to
$\true$. Finally, process constants provide a mechanism for
including recursion in the calculus.

Moving on to the action of reproduction, to capture the
creation of new individuals, we employ the special \emph{species}
processes $R$. $R$, defined as $!rep.P$, are replicated processes which
may continuously receive input through channel $rep$ and create
new instances of process $P$, where $P$ is a new individual of
species $R$. Such inputs will be provided
by individuals in the phase of reproduction via the complementary action $\ol{rep}$.

Finally, population systems are built by composing in parallel located
individuals, $P\loc{\s,\ell}$, where $\s$ and $\ell$ are the species
and the location of the individual, and species $R\loc{\s}$, where
$\s$ is the name of the species. Finally, $S\hide L$ models the
restriction of the use of channels in set $L$ within $S$.

As an example, we consider the model described in ~\cite{BS05} where a
set of individuals live on an $n\times n$ lattice of resource sites
and go through phases of reproduction and dispersal. Specifically,
the studied model considers a population where individuals disperse
in space while competing for a location site during their
reproduction phase. They produce an offspring only if they have
exclusive use of a location. After reproduction the offspring
disperse and continue indefinitely with the same behavior. In PALPS,
we may model the described species $\s$ as $R\eqdef !rep.P$, where
\begin{eqnarray*}
\vspace{0.1in}
P &\eqdef &\sum_{\ell\in Neigh(\myloc)}\frac{1}{4}: go\,\ell.\surd.\cond(\s@\myloc = 1 \rhd P_1;\;\true \rhd \surd.P)\\
P_1 &\eqdef & p:\ol{rep}.\surd. P_1 + (1-p):\ol{rep}.\ol{rep}.\surd. P_1
\vspace{0.1in}
\end{eqnarray*}
We  point out that the conditional construct allows us to determine
the exclusive use of a location by an individual. The special label
$\myloc$ is used to illustrate that the location of interest is the
actual location of an individual once the individual is placed in a
context within a system definition.
% where $\els$
%is used as a shorthand to $\lnot(\s@\ell = 1)$.
Furthermore, note that $P_1$ models the probabilistic production
of one or two offsprings of the species. During the dispersal phase,
an individual moves to a neighboring location which is chosen
probabilistically among the four neighboring locations on
the lattice of the individual. Then a system containing of two individuals at a location
$\ell$ and one in location $\ell'$ can be modeled as
\vspace{0.1in}
\[System \eqdef (P\loc{\ell,\s} |P\loc{\ell,\s} |P\loc{\ell',\s}|
(!rep.P)\loc{\s})\hide\{rep\}.\]
\vspace{0.1in}
To model a competing species $\s'$ which preys on $\s$, we may
define the process $R'\eqdef !rep'.Q$, where
\vspace{0.1in}
\begin{eqnarray*}
Q &\eqdef & \cond(\s@\myloc > 1 \rhd \ol{prey_s}.\surd. Q_1, \true\rhd \surd.Q_2)\\
Q_1 & \eqdef & \ol{rep'}.\surd.Q\\
Q_2 & \eqdef &  \cond(\s@\myloc > 1 \rhd \ol{prey_s}.\surd. Q_1, \true\rhd \surd.\nil)
\end{eqnarray*}
An individual of this species looks for a prey. If it succeeds in locating one, then it produces an
offspring. If it fails for two consecutive time units it dies.

\subsection{The Semantics}

The semantics of PALPS is defined in terms of a structural
operational semantics given at the level of configurations of the form
$(E,S)$, where $E$ is an \emph{environment} and $S$ is a population system. The environment
$E$ is an entity which captures how the various locations of the system are populated.
More precisely, $E\subset \L\times \S\times \mathbb{N}$, where each pair $\ell$ and $\s$ is represented
in $E$ at most once and where
$(\ell, s,m)\in E$ denotes the existence of $m$ individuals of
species $s$ at location $\ell$.
The environment $E$ plays a central role in defining
the semantics of the calculus and, in particular, for evaluating  expressions.
The satisfaction relation for logical expressions $\models$ is defined inductively on the
structure of a logical expression as shown in Table~\ref{evals}.

\begin{table}[h!]
\caption{\label{evals}{\bf The satisfaction relation for logical expressions}}
\begin{center} {\small
\begin{tabular}{l l l l }\hline\\
     \vspace{0.05in}
$E\sat \true$ & always\\     \vspace{0.05in}
$E \models \lnot e$ & if and only if &
$\lnot(
E\models e)$\\
     \vspace{0.05in}
$E\models e_1\land e_2$ & if and only if & $E\models e_1 \land
E\models e_2$\\
     \vspace{0.05in}
$E\models w\bowtie e$ & if and only if & $\val(E,w) \bowtie e$\\
 \hline
\end{tabular}}
\end{center}
\end{table}

The relation $\models$ is straightforward and depends on the
evaluation function for arithmetic expressions $\val(E,w)$ defined
in Table~\ref{models}.

\begin{table}[h!]
\caption{\label{models}{\bf The evaluation relation for arithmetic expressions}}
\begin{center} {\small
\begin{tabular}{l l l l }\hline\\
$\val(E,c)$ & $=$ & $c$\\
$\val(E,\psi@{\ell})$ & $=$ & $\psi_{\ell}$\\
$\val(E,\s@\ell)$ & $=$ & $\num(E,\ell,\s)$\\
$\val(E,@\ell)$ & $=$ & $\num'(E,\ell)$\\
$\val(E,\op_1(w))$ & $=$ & $\op_1(\val(E,w))$\\
$\val(E,\op_2(w_1,w_2))$ & $=$ & $\op_2(\val(E,w_1),\val(E,w_2))$\\
\\
\hline
\end{tabular}}
\end{center}
\end{table}

The auxiliary functions $\num(E,\ell,\s)$ and $\num'(E,\ell)$
compute the number of individuals at location $\ell$ in environment $E$
of a specific species $\s$ ($\num(E,\ell,\s)$) or for all species
($\num'(E,\ell)$) and are defined by $\num(E,\ell,\s)= n$ where $(\ell,\s,n)\in E$ and
$\num'(E,\ell)= \sum_{s\in S} \num(E,s,\ell)$.

Before we proceed to the semantics we define some additional operations on environments that we will use in the sequel:

\begin{definition}
Consider environment $E$ location $\ell$ and species  $\s$.
\begin{itemize}
\item $E\oplus {(\s,\ell)}$ increases the count of individuals of species $\s$ at location $\ell$ in environment $E$ by $1$:
\begin{equation*}
  E\oplus {(\s,\ell)}=  \left\{
  \begin{array}{ll}
    E'\cup \{(\ell, \s,m+1)\} & \mbox{ if } E= E'\cup \{(\ell, \s,m)\} \mbox { for some } m\\
    E \cup \{(\ell, \s,1)\} & \mbox{ otherwise}
  \end{array} \right.
\end{equation*}
\item $E\ominus {(\s,\ell)}$ decreases the count of individuals of species $\s$ at location $\ell$ in environment $E$ by $1$:\begin{equation*}
  E\ominus {(\s,\ell)}=  \left\{
  \begin{array}{ll}
    E'\cup \{(\ell, \s,m-1)\}& \mbox{ if } E=E'\cup \{(\ell, \s,m)\}, m>1\\
     E' & \mbox{ if }E=E'\cup \{(\ell, \s,1)\}\\
      \bot & \mbox{ otherwise }
  \end{array} \right.
\end{equation*}
\end{itemize}
\end{definition}
We may now define the semantics of PALPS, presented in
Tables~\ref{op-sem-nodes} and~\ref{op-sem-networks}, and given in
terms of two transition relations, the nondeterministic relation
$\nby{}$ and the probabilistic relation $\pby{}$. A transition of
the form $(E,S)\nby{\mu}(E',S')$ signifies that configuration
$(E,S)$ may execute action $\mu$ and become $(E',S')$ whereas a
transition of the form $(E,S)\pby{w}(E',S')$ signifies that
configuration $(E,S)$ may evolve into configuration $(E',S')$ with
probability $w$. Whenever the type of the transition is irrelevant
to the context, we write $(E,S)\by{\a} (E',S')$ to denote that either
$(E,S)\nby{\mu}(E',S')$  or $(E,S)\pby{w}(E',S')$. Action $\mu$
appearing in the nondeterministic relation may have one of the
following forms:

\begin{itemize}
 \item $a@\ell$ and $\ol{a}@\ell$ denote the execution of actions $a$ and $\ol{a}$
 respectively at location $\ell$.
% \item $prey_{\s}@\ell$ denotes the execution of a prey action at
% location $\ell$ by an individual belonging to the species $\s$.
 \item $\tau$ denotes the internal action. This may arise when two
 complementary actions take place at the same location or when a move or a prey action
 take place. We are not interested in the precise location of
 internal actions, thus, this information is not included.
 \item $\surd$ denotes the time passing action.
\end{itemize}

\begin{table*}[h!]
\caption{\label{op-sem-nodes}{\bf Transition rules for individuals}}
\begin{center} {\small
\begin{tabular}{l l l l }\hline
        \vspace{0.1in}
\axiom{Tick}&
 $(E,\surd.P\loc{s,\ell})\nby{\surd}(E^P,P\loc{s,\ell})$ \\ %&\hspace{1cm}
        \vspace{0.1in}
        \axiom{Act}& %\hspace{-1,7cm}
 $(E,\eta.P\loc{s,\ell}\nby{\eta@\ell}(E^P,P\loc{s,\ell})
 $ &$\eta \neq go\,\ell'$
\\       \vspace{0.1in}
\axiom{Go}& %\hspace{-1,7cm}
 $(E,go\, \ell'.P\loc{s,\ell})\nby{\tau}((E\ominus(s,\ell))\oplus(s,\ell'),P\loc{s,\ell'})
 $ & $(\ell,\ell')\in\Nb$
\\ \vspace{0.2in}
\axiom{Prey}&
 $(E,P\loc{s,\ell})\nby{prey_s@\ell}(E\ominus(s,\ell), \nil\loc{s,\ell})$ \\ \vspace{0.2in}%&\hspace{1cm}

\axiom{PSum}&
    $ {(E,\sum_{i\in I} w_i:P_i\loc{s,\ell}) \pby{\val(E,w_i\!\downarrow\!\ell)} (E^{P_i},P_i\loc{s,\ell}})$\\ \vspace{0.1in}
%  &\hspace{1cm}
\axiom{Const}& %\hspace{-0,7cm}
    $\fracc{(E,P\loc{s,\ell}) \by{\alpha}(E',P'\loc{s,\ell})} {(E,C\loc{s,\ell})\by{\a}(E',P'\loc{s,\ell})}$
    $\;\;\;C\eqdef P\loc{s,\ell}$ \\ \vspace{0.3in}
\axiom{Cond}& %\hspace{-1,3cm}
$\fracc{(E,P_i\loc{s,\ell})\by{\alpha}(E',P_i'\loc{s,\ell'}), E\sat e_i\!\downarrow\!\ell, E\not\sat e_j\!\downarrow\!\ell, j<i} {(E,\cond(e_1\rhd
P_1,\ldots, e_n\rhd P_n))\by{\a}(E',P_i'\loc{s,\ell'})}$\\
\vspace{0.4cm}
& \hspace{0.5in}where
  $E^P=  \left\{
  \begin{array}{ll}
    E\ominus{(\s,\ell)}& \mbox{ if } P = \nil\\
    E & \mbox{ otherwise}
  \end{array} \right.$\\
\hline %\vspace{-0.2in}
\end{tabular}}
\end{center}
\end{table*}

The rules of Table~\ref{op-sem-nodes} prescribe the semantics of
located individuals in isolation. The first four axioms define
nondeterministic transitions, the fifth axiom defines a
probabilistic transition, and the last two rules refer to both the
nondeterministic and the probabilistic case. All rules are concerned
with the evolution of the individual in question and the effect of
this evolution to the system's environment. A key issue in the
enunciation of the rules is to preserve the compatibility of $P$ and
$E$ as transitions are executed. We consider each of the rules
separately. Axiom $\axiom{Tick}$ specifies that a $\surd$-prefixed
process will execute the time consuming action $\surd$ and then
proceed as $P$. The state of the new environment depends on the
state of $P$: if $P= \nil$ then the individual has terminated its
computation and, therefore, it is removed from $E$ (see the
definition of $E^P$) whereas, if $P\neq \nil$ then, obviously, $E$
remains unchanged. Axiom $\axiom{Act}$ specifies that $\eta.P$
executes action $\eta@\ell$ and evolves to $P$. Note that the action
is decorated by the location of the individual executing the
transition to enable synchronization of the action with
complementary actions taking place at the same location (see rule
\axiom{Par2}, Table~\ref{op-sem-networks}). This axiom excludes the
case of $\eta = go\,\ell$ which is treated separately in the next
axiom. Specifically, according to Axiom $\axiom{Go}$, an individual
may change its location. This gives rise to action $\tau$ and has
the expected effect on the environment $E$. Moving on to Axiom
$\axiom{Prey}$, this describes that any individual can become the
victim of a preying action. This may happen at any point during the
lifetime of the individual giving rise to the action
$prey_{\s}@\ell$ and causing the individual to terminate with the
appropriate changes to the state of the environment.  Rule
$\axiom{PSum}$ expresses the semantics of probabilistic choice: once
the probability expressions are evaluated within the environment,
the probabilistic action is taken leading to the appropriate
continuation: if the resulting state of the individual, namely
$P_i$, is equal to $\nil$, then the individual is removed from the
environment $E$. Note that we write $w\!\downarrow\!\ell$ for the expression $w$
with all occurrences of $\myloc$ substituted by location $\ell$:
$w\!\downarrow\!\ell = w[\ell/\myloc]$. Next $\axiom{Const}$ express the
semantics of process constants in the expected way. Finally, rule
$\axiom{Cond}$ stipulates that a conditional process may perform an
action of continuation $P_i$ assuming that $e_i\!\downarrow\!\ell$ evaluates to
true and all $e_j\!\downarrow\!\ell$, $j<i$ evaluate to false. Similarly to
$w\!\downarrow\!\ell$,  $e\!\downarrow\!\ell$ is the expression $e$ with all occurrences of
$\myloc$ substituted by location $\ell$.

We may now move on to Table~\ref{op-sem-networks} which defines the
semantics of system-level operators. The first rule defines the
semantics for the replication operator, the next five rules define
the semantics of the parallel composition operator, and the last
rule deals with the restriction relation.

\begin{table*}[h!]
\caption{\label{op-sem-networks}{\bf Transition rules for systems}}
\begin{center} {\small
\begin{tabular}{l l l l }\hline\\
     \vspace{0.14in}
\axiom{Rep}&%\hspace{-1in}
$\fracc{R = !rep_\s.P\loc{\s},\,\ell\in\L}
{(E,R)\nby{rep_\s@\ell} (E\oplus (\s,\ell), P\loc{\s,\ell}| R)}$\\
 \vspace{0.12in}

\axiom{Par1}&
    $\fracc{(E,S_1)\nby{\mu} (E',S_1'), (E,S_2)\not\!\!\pby{}}
    {(E,S_1 | S_2)\nby{\mu} (E',S_1' | S_2)}$\\
 \vspace{0.12in}
    \axiom{Par2}&%\hspace{-1in}
 $\fracc{(E,S_1)\nby{a@\ell} (E_1,S_1'), (E,S_2)\nby{\ol{a}@\ell}(E_2,S_2')}
 {(E,S_1| S_2)\nby{\tau} (E\otimes(E_1,E_2),S_1' | S_2')}$\\
 \vspace{0.12in}
 \axiom{Par3}&
    $\fracc{(E,S_1)\pby{w_1} (E_1,S_1'), (E,S_2)\pby{w_2}(E_2,S_2')}
 {(E,S_1| S_2)\pby{w_1\cdot w_2} (E\otimes(E_1,E_2),S_1' | S_2')}$\\ \vspace{0.12in}
    %\hspace{0.3in}
    \axiom{Par4}&%\hspace{-1in}
 $\fracc{(E,S_1)\pby{w} (E',S_1'), (E,S_2)\not\!\!\pby{}}
 {(E,S_1| S_2)\pby{w} (E',S_1' | S_2)}$\\
 \vspace{0.12in}
 \axiom{Time}&
 $\fracc{(E,S_1)\nby{\surd} (E_1,S_1'), (E,S_2)\nby{\surd}(E_2,S_2')}
 {(E,S_1| S_2)\nby{\surd} (E,S_1' | S_2')}$\\
 \vspace{0.12in}
 \axiom{Res}&
 $\fracc{(E,S)\by{\a} (E',S'), \a \not\in\{a@\ell,\ol{a}@\ell | a\in L\}}
 {(E,S \hide L) \by{\a} (E',S')\hide L }$\\
\hline \vspace{-0.4in}
\end{tabular}}
\end{center}
\end{table*}
%\vspace{-0.2in}

Thus, according to axiom $\axiom{Rep}$, a species process may execute
action $rep_\s@\ell$ for any location $\ell$ %. In this way it may
%synchronize with any complementary $\ol{rep}_s@\ell$
and create
a new individual $P$ of species $\s$ at location $\ell$.
 Next, rules $\axiom{Par1}$ - $\axiom{Par4}$  specify
how the actions of the components of a parallel composition may be combined.
Note that the symmetric versions of these rules are omitted.
According to $\axiom{Par1}$, if a component may execute a nondeterministic transition
and no probabilistic transition is enabled by the other component (denoted by $(E,S_2)\not\!\!\pby{}$), then the transition may take place.
If the parallel components may execute complementary actions, then they
may synchronize with each other producing action $\tau$ (rule $\axiom{Par2}$).
If both components may execute probabilistic transitions then they may
proceed together with probability the product of the two distinct
probabilities (rule $\axiom{Par3}$) and, finally, if exactly one of them
enables a probabilistic transition then this transition takes precedence
over any nondeterministic transitions of the other component (rule $\axiom{Par4}$).
Note that in case that the components proceed simultaneously then
the environment of the resulting configuration should take into account
the changes applied in both of the constituent transitions
(rules $\axiom{Par2}$ and $\axiom{Par4}$. This is implemented by $E\otimes(E_1,E_2)$ as follows:
\[ E\otimes (E_1, E_2)  =  \{(\ell, \s, m+i_1+i_2) \mid (\ell, \s, m) \in E,  (\ell, \s, m+i_1)\in E_1, (\ell, \s, m+i_2)\in E_2, i_1,i_2\in\mathbb{Z}\} \]
%\begin{eqnarray*}
% E\otimes (E_1, E_2) & = & \{(\ell, \s, m) \mid (\ell, \s, m) \in E \cap E_1\cap E_2\}\\
%            &\cup & \{(\ell, \s, m) \mid (\ell, \s, m)  \in E , (\ell, \s, m-1)  \in  E_1, (\ell, \s, m+1)  \in  E_2\}\\
%            &\cup & \{(\ell, \s, m-1) \mid (\ell, \s, m)  \in E\cap E_i , (\ell, \s, m-1)  \in  E_{3-i}, i\in\{1,2\}\}\\
%            &\cup & \{(\ell, \s, m-2) \mid (\ell, \s, m)  \in E, (\ell, \s, m-1)  \in  E_1\cap E_2\}\\
%            &\cup & \{(\ell, \s, m+1) \mid (\ell, \s, m)  \in E\cap E_i , (\ell, \s, m+1)  \in  E_{3-i}, i\in\{1,2\}\}
%\end{eqnarray*}
Next, rule $\axiom{Time}$ defines that parallel processes must synchronize on $\surd$ actions, thus allowing one tick of time to pass and all processes to proceed to their next round. Finally, rule $\axiom{Res}$ defines the semantics of the restriction operator in the usual way.

Based on this machinery, the semantics of a system $S$ is obtained by applying the semantical rules to the initial configuration. The initial configuration, $(E,S)$, is such that  $(\ell, \s, m)\in E$ if and only if $S$ contains exactly $m$ individuals of species $\s$ located at $\ell$. In general, we say that $E$ is \emph{compatible }with $S$ whenever $(\ell, \s, m)\in E$ if any only if $S$ contains exactly $m$ individuals of species $\s$ located at $\ell$. It is possible to prove the following lemma by structural induction on $S$~\cite{Mar-thesis}.
\begin{lemma}
Whenever $(E,S)\by{\a}(E',S')$ and $E$ is compatible with $S$, then $E'$ is also compatible with $S'$.
\end{lemma}

\subsection{Model Checking PALPS}

Model-checking of PALPS processes may be implemented via
an instantiation of
the PCTL logic~\cite{CTL1}. The instantiation involves the adoption of PALPS logical expressions as
the atomic propositions of the logic. Specifically,
the syntax of the PCTL instantiation that we consider, is given by the following grammar where $\Phi$ and $\phi$ range over PCTL state and path formulas, respectively, $p\in[0,1]$ and $k\in \mathbb{N}$.
\begin{eqnarray*}
\Phi&:=&true\;\;\mid\;\;e\;\;\mid\;\;\lnot \Phi\;\;\mid\;\;\Phi\land \Phi'\;\;\mid\;\; {\sf P}_{\bowtie p} [\phi]\\
\phi & := & {\sf X}\Phi \;\;\mid\;\; \Phi {\sf U}^k \Phi \;\;\mid\;\; \Phi_1 {\sf U} \Phi
\end{eqnarray*}

In the syntax above, we distinguish between state formulas $\Phi$ and path formulas
$\phi$, which are evaluated over states and paths, respectively. A state formula is built
over PALPS logical expressions and the construct ${\sf P}_{\bowtie p} [\phi]$. Intuitively, a configuration
$s$ satisfies property ${\sf P}_{\bowtie p} [\phi]$ if for any possible execution beginning at
the configuration, the probability of taking a path that satisfies the path formula $\phi$ satisfies
the condition $\bowtie p$.
Path formulas include the ${\sf X}$ (next), ${\sf U}^k$ (bounded until) and ${\sf U}$ (until)
operators, which are standard in temporal logics. Intuitively, ${\sf X} \Phi$ is satisfied
in a path if the next state satisfies path formula $\Phi$,  $\Phi_1 {\sf U}^k \Phi_2$ is
satisfied in a path if $\Phi_1$ is satisfied continuously on the path until $\Phi_2$ becomes true
within $k$ time units (where time units are measured by $\surd$ events in PALPS) and
$\Phi_1 {\sf U} \Phi_2$ is satisfied if $\Phi_2$ is satisfied at some
point in the future and $\Phi_1$ holds up until then.

For example, consider a population $\s$ in danger of extinction. A
property that one might want to check for such a population is that the probability
of extinction of the population in the next ten years is less than a certain threshold $p_e$. This can be expressed
in PCTL by the property ${\sf P}_{\leq p_e}[true {\sf U}^{10} \sum_{\ell\in \L} \s@\ell = 0]$. Alternatively,
one might express that a certain central location $\ell$ will be  reinhabited with at least some probability
$p_r$ by: $\s@\ell = 0 \rightarrow {\sf P}_{\geq p_r}[true{\sf U} (\s@\ell > 0)]$. Similarly, it would
 be possible to study the relation within a model between the size of the initial population and the probability of extinction of the population,
 by checking properties
of the form $\s@\ell \geq m \rightarrow {\sf P}_{\geq p_e}[true{\sf U} (\s@\ell = 0)]$ or explore the
dynamics between two (or more) competing populations $\s$ and $\s'$ by, for example, expressing that within
 the next 20 years with some high probability, members of the population $\s$ will outnumber the
members of population $\s'$: ${\sf P}_{\geq p} [true {\sf U} (\sum_{\ell\in \L} \s'@\ell \leq \sum_{\ell\in \L} \s@\ell)]$.

The semantics of PCTL are defined over Markov Decision Processes (MDPs), a type of transition systems that combine
probabilistic and nondeterministic behavior. It is not difficult to see that the operational semantics of PALPS
gives rise to transition systems that can easily be translated to MDPs~\cite{Mar-thesis}. For the details of the
semantics and the model checking algorithm we refer the reader to~\cite{CTL2}.

As a final note we observe that in order to check the satisfaction
of PCTL properties by  PALPS processes it is sufficient to restrict
our attention to the $E$ component of each configuration $(E,S)$.
This is due to the fact that $E$ is the only information required in
order decide the satisfaction of logical expressions by
configurations (see Tables~\ref{evals} and~\ref{models}).

\section{Examples}\label{exampl}

During the last few decades, the theory of metapopulations has been an active field of research in Ecology and it has been extensively studied by conservation scientists and landscape ecologists to analyze the behavior of interacting populations and to determine how the topology of fragmented habitats may influence various aspects of these systems such as local and global population persistence and species evolution.
The notion of a metapopulation refers to a group of distinct populations of the same species residing on a fragmented habitat or, a so-called set of patches, and cycle in relative independence through their life cycle while interacting with other populations and colonizing previously unoccupied locations through dispersal. It has been observed that while populations of a metapopulation may go extinct as a consequence of demographic stochasticity, the metapopulation as a whole is often stable because immigrants from another population are likely to re-colonize habitat which has been left open by the extinction of another population or because immigration to a small population may rescue that population from extinction.  Indeed the process of dispersal is of vital importance in metapopulations. It affects the long-term persistence of populations, the coexistence of species and genetic differentiation between subpopulations and understanding this process is essential for obtaining a good understanding of the behavior of metapopulations. The evolution of dispersal has received much attention by scientists and it has been studied in connection to various parameters such as the connectivity of the habitat on which a metapopulation exists, patch quality and local dynamics.

In this section, we describe two examples relating to metapopulation
dispersal through which we illustrate how our calculus can be used to construct
models of this phenomenon.

\begin{example}{\rm\
The first example we consider is motivated by the spatially-explicit, individual-based
model of~\cite{Travis-Dytham}. In this work the authors construct a fairly simple model of
metapopulation dispersal which departs from previous works in that, unlike
previous models of metapopulation dispersal which tended to be deterministic and at the level of population densities,
the model constructed is both stochastic and individual-based.

\begin{figure}[t]
  \centerline{\includegraphics[width=.7\textwidth]{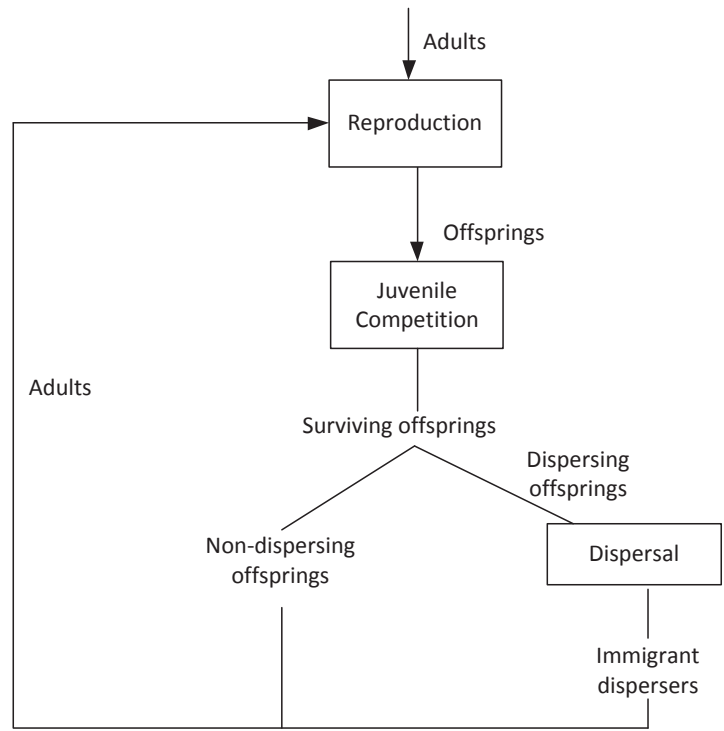}}
  \caption{The sequence of events in the lifetime of a dispersing species}
  \label{fig:ex1}
\end{figure}

According to this study, a set of genotypes co-exist within a
habitat which differ only in their propensity to disperse. The
metapopulation is composed of $n\times n$ subpopulations inhabiting
a set of patches arranged on a square lattice with cyclic
boundaries, so that individuals leaving the ``top'' or
``right-side'' of the world reappear on the ``bottom'' or
``left-side'' respectively and vice versa. Each patch is associated
with a so-called patch quality related to the capacity of the patch.
The behavior of an individual of the genotypes under study is
illustrated diagrammatically in Figure~\ref{fig:ex1}. According to
this model, an adult individual initially produces $\lambda$
offsprings. Subsequently, a phase of competition takes place between
the juveniles of the population of which a fraction survives. Each
surviving offspring may disperse according to a probability of
dispersal distinct to its genotype. In case it disperses, the
neighboring patch it moves to is selected with equal probability
among all neighbors. This sequence of events in the behavior of an
individual is presented diagrammatically in Figure~\ref{fig:ex1}. We
point out that the percentage of offspring surviving juvenile
competition at patch $\ell$ is given by $\gamma_{\ell} =
(1+\alpha_{\ell}\cdot N_{\ell})^\beta$, where $\alpha_{\ell}$ is the
measure of the patch quality, $N_{\ell}$ is the number of
individuals residing at patch $\ell$ and $\beta$ is a constant that
relates to the degree of competition.

This metapopulation can be modeled in PALPS as follows. We consider the set of of locations $(i,j)$, $1\leq i, j\leq n$, where two locations
$(i,j)$ and $(k,l)$ are neighbors if they are adjacent on the grid. Finally, let us consider the location attribute $\a_{\ell}$ as a measure of the quality of the patch at $\ell$. Then, genotype $i$ with some constant probability of dispersal $p_i$ and $\lambda = 3$ can be defined as
the species process $R_i = !rep_i.J_i$, where
\begin{center}
\begin{tabular}
{l l l l }
$A_i$ & $\eqdef$ & $\ol{rep}_i.\ol{rep}_i.\ol{rep}_i.0$ & \hspace{1in}Adult Individual\\
$J_i$ & $\eqdef$ & $q_i:S_i + (1-q_i):\nil$ &\hspace{1in}Juvenile\\
$S_i$ & $\eqdef$ & $p_i :D_i + (1-p_i) : \surd.A_i$ & \hspace{1in}Surviving Juvenile\\
$D_i$ & $\eqdef$ & $\sum_{\ell\in Neigh(\myloc)}\frac{1}{4}: go\,\ell.\surd.A_i$ &\hspace{1in}Dispersing Juvenile
\end{tabular}
\end{center}
and $q_i$ the probability of survival of juvenile competition is given by $q_i = (1+\alpha_{\ell}\cdot @\ell)^\beta$.
Then a system can be modeled as the composition of the various genotypes as well as the individuals of the initial population under study:
\[System \eqdef [(R_1\loc{1} \mid \ldots R_k\loc{k}\mid  \prod_{1\leq i \leq m_1} A_1\loc{\ell_1,1} \mid \ldots)\hide \{rep_1,\ldots rep_k\}.\]

Analysis in this model may focus on the effect that the dispersal rates, the degree of competition and/or patch quality may have on the degree of population dispersals.
}\end{example}

\begin{example}{\rm\
As another more complex example, let us consider a model of wood
thrush dispersal, initially proposed in~\cite{wp-one} and expanded
upon in~\cite{wp-two}. This model considers three types of birds:
adult breeders, adult floaters, and juveniles which are birds in
their first year of life. According to this model, adult breeders
produce an offspring at a rate dictated by various system parameters
such as clutch size, nest predation and paratisism rates which we
denote as $r_b$. Following reproduction, each individual has a
probability of dying before the next time step which is higher in
juveniles and adult floaters in comparison to adult breeders. We
write $q_b$, $q_j$ and $q_f$ for the mortality rates of breeders,
juveniles and floaters, respectively. If following mortality a
habitat patch has more birds than its capacity allows, then
dispersal will occur according to a probability determined by the
size of the patch and the distance between neighboring patches. This
probability is higher in floaters and juveniles in comparison to
adult breeders who exhibit a high site fidelity. We write $p_b$,
$p_j$ and $p_f$ for the dispersion rates of breeders, juveniles and
floaters, respectively.  If a bird reaches a patch with available
capacity then it will settle. If not, then it will either attempt to
disperse to another patch or it will become a floater depending on
whether it has reached its maximum number of dispersal events. Once
dispersal has occurred, the juveniles become adults and the model
begins another cycle.  This sequence of events in the behavior of
the populations is presented diagrammatically in
Figure~\ref{fig:ex2}.

\begin{figure}
  \centerline{\includegraphics[width=.8\textwidth]{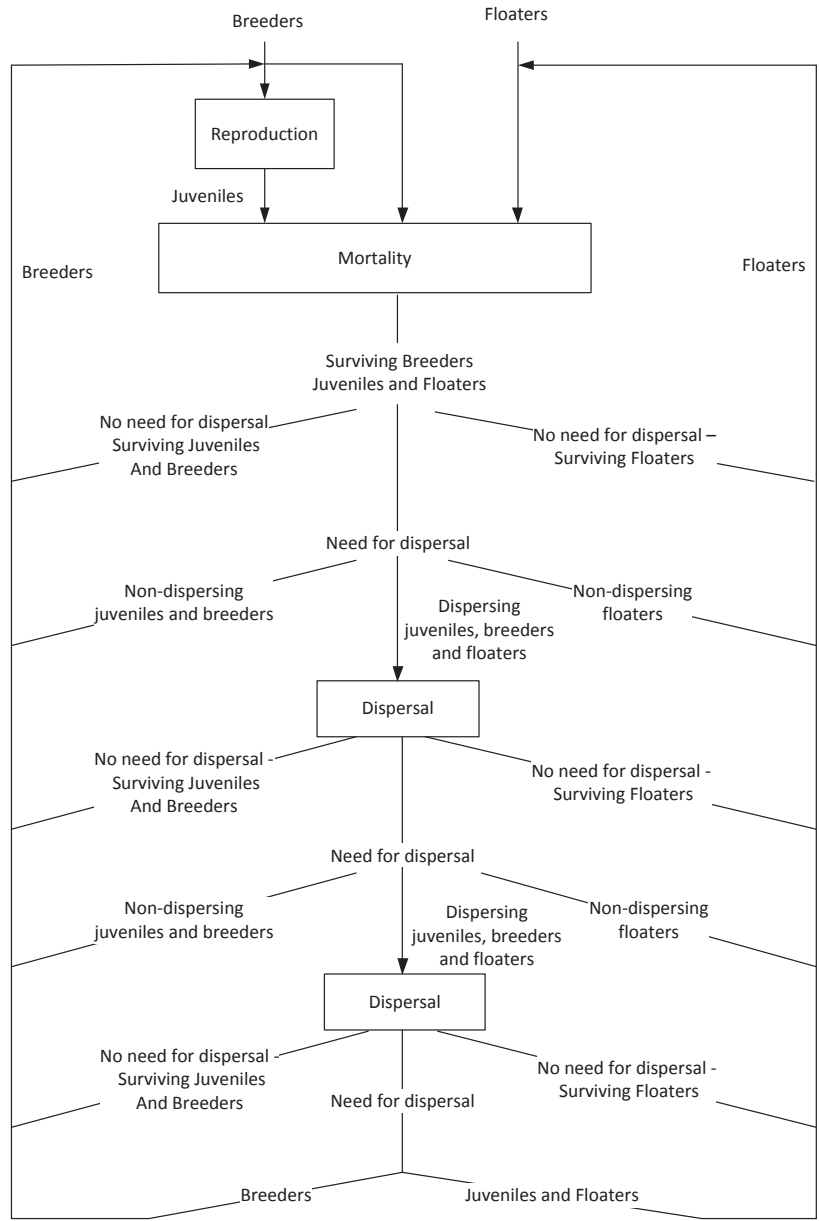}}
  \caption{A cycle in the lifetime of the metapopulation}
  \label{fig:ex2}
\end{figure}

This metapopulation can be modeled in PALPS as follows. We consider the set of of locations and an associated predefined neighbor function as well as a distance function that may be instantiated according to the modeler's preference to capture Euclidean distance or some other function of interest~\cite{wp-two}. We also assume the existence of a set of probabilities $\{p_{i,j}\}_{i,j\in\L}$ where $p_{i,j}$ represents the probability of dispersal from patch $i$ to patch $j$. Finally, we introduce the location attribute $c_{\ell}$ as a measure of the capacity of patch $\ell$.
Then, wood thrush species can be modeled by the process $R = !rep.Juv$, where the behavior of a juvenile individual $J_i$ is described by the following equations:
% $i$ with probability of dispersal $p_i$ and $\lambda = 3$ can be defined as
%the species process $R_i = !rep_i.J_i$, where
%
\begin{center}
\begin{tabular}{r l l r }
$Juv$ & $\eqdef$ & $q_j:JC_0 + (1-q_j):\nil$ &Juvenile survival\\
$JC_0$ & $\eqdef$ & $\cond(@\myloc > c_{\myloc}\rhd JD_0,\true\rhd\surd.AB)$ &Check patch capacity\\
$JD_0 $& $\eqdef$ & $p_j:JA_1 + (1-p_j):\surd.AB$ &Decide whether to disperse\\
$JA_1$ & $\eqdef$ &  $\sum_{\ell\in Neigh(\myloc)}p_{\myloc,\ell}: go\,\ell.JC_1$&Dispersal attempt $1$\\
$JC_1$ & $\eqdef$ & $\cond(@\myloc > c_{\myloc}\rhd JD_1,\true\rhd\surd.AB)$ &Check patch capacity\\
$JD_1 $& $\eqdef$ & $p_j:JA_2 + (1-p_j):\surd.AB$ &Decide whether to disperse\\
$JA_2$ & $\eqdef$ &  $\sum_{\ell\in Neigh(\myloc)}p_{\myloc,\ell}: go\,\ell.JC_2$&Dispersal attempt $2$\\\vspace{0.15in}
$JC_2$ & $\eqdef$ & $\cond(@\myloc > c_{\myloc}\rhd \surd.Fl,\true\rhd\surd.AB)$&Become floater or adult\\
$AB$ &$\eqdef$ & $r_b: \ol{r_b}_i.BS + (1-rb):BS$ &Breeder reproduction\\
$BS$ & $\eqdef$ & $q_b:BC_0 + (1-q_b):\nil$ &Breeder survival\\
$BC_0$ & $\eqdef$ & $\cond(@\myloc > c_{\myloc}\rhd BD_0,\true\rhd\surd.AB)$ &Check patch capacity\\
$BD_0 $& $\eqdef$ & $p_b:BA_1 + (1-p_b):\surd.AB$ &Decide whether to disperse\\
$BA_1$ & $\eqdef$ &  $\sum_{\ell\in Neigh(\myloc)}p_{\myloc,\ell}: go\,\ell.BC_1$&Dispersal attempt $1$\\
$BC_1$ & $\eqdef$ & $\cond(@\myloc > c_{\myloc}\rhd BD_1,\true\rhd\surd.AB)$ &Check patch capacity\\
$BD_1 $& $\eqdef$ & $p_b:BA_2 + (1-p_b):\surd.AB$ &Decide whether to disperse\\
$BA_2$ & $\eqdef$ &  $\sum_{\ell\in Neigh(\myloc)}p_{\myloc,\ell}: go\,\ell.BC_2$&Dispersal attempt $2$\\\vspace{0.15in}
$BC_2$ & $\eqdef$ & $\cond(@\myloc > c_{\myloc}\rhd \surd.Fl,\true\rhd\surd.AB)$&Floater or adult\\
$Fl$ & $\eqdef$ & $ q_f:FC_0 + (1-q_f):\nil$&Floater survival\\
$FC_0$ & $\eqdef$ & $\cond(@\myloc > c_{\myloc}\rhd FD_0,\true\rhd\surd.Fl)$ &Check patch capacity\\
$FD_0 $& $\eqdef$ & $p_f:FA_1 + (1-p_f):\surd.Fl$ &Decide whether to disperse\\
$FA_1$ & $\eqdef$ &  $\sum_{\ell\in Neigh(\myloc)}p_{\myloc,\ell}: go\,\ell.FC_1$&Dispersal attempt $1$\\
$FC_1$ & $\eqdef$ & $\cond(@\myloc > c_{\myloc}\rhd FD_1,\true\rhd\surd.Fl)$ &Check patch capacity\\
$FD_1 $& $\eqdef$ & $p_f:FA_2 + (1-p_f):\surd.Fl$ &Decide whether to disperse\\
$FA_2$ & $\eqdef$ &  $\sum_{\ell\in Neigh(\myloc)}p_{\myloc,\ell}: go\,\ell.\surd.Fl$&Dispersal attempt $2$
\end{tabular}
\end{center}
As before, the system can be modeled as the composition of the species as well as the various individuals that form the study:
\[System \eqdef [(R\loc{1}  \mid \prod_{1\leq i\leq n_b^1} AB\loc{\ell_1,1} \mid \mid \prod_{1\leq i\leq n_j^1} Juv\loc{\ell_1,1}\mid \prod_{1\leq i\leq n_f^1} Fl\loc{\ell_1,1}\ldots)\hide \{rep_1,\ldots rep_k\}.\]
Varying the model parameters, e.g. the habitat topology, patch quality and dispersal distance, may allow an analysis of the effects of the parameters on patch and metapopulation persistence.
}\end{example}

\section{Concluding remarks}\label{remarks}
This paper reports on work towards the development of a
process-calculus framework for the spatially-explicit and individual-based
modeling of ecological systems. In related work~\cite{Mar-thesis}
we have also implemented a prototype tool and conducted simulations for
the spatially-explicit model of~\cite{BS05}. In future work we intend to provide
optimizations for our tool via an implementation of a spatial extension of
the Gillespie simulation algorithm~\cite{CPBM06,JEU-11}
and by taking advantage of concepts developed in process-algebraic
frameworks for state-space reduction such as confluence and minimization according
to equivalence relations. At the same time
it is our intention to enhance the syntax of PALPS to enable a more succinct
presentation of systems especially in terms of the multiplicity of individuals.
Other possible directions for future work include the adoption of continuous
time as well as the use of dynamic attributes to allow
exploring the system while patch quality degrades, temperatures increase, etc.

\bibliographystyle{eptcs}
\bibliography{mecbic}

\end{document}